# Upcycling Low-Nickel Polycrystalline Cathodes from Retired Electric Vehicle Batteries into Single-Crystal Nickel-Rich Cathodes

Guannan Qian,[1,2,#] Zhiyuan Li,[1,#] Yong Wang,[1] Xianyu Xie,[3] Yushi He,[1] Jizhou Li,[2] Yanhua Zhu,[4] Zhengjie Chen,[5] Sijie Xie,[5] Haiying Che,[1] Yanbin Shen,[5] Liwei Chen,[5,6] Xiaojing Huang,[7] Zi-Feng Ma,[1] Yijin Liu,[2,*] Linsen Li[1,8,*]

[1] Department of Chemical Engineering, Shanghai Electrochemical Energy Device Research Center (SEED), School of Chemistry and Chemical Engineering, Frontiers Science Center for Transformative Molecules, Shanghai Jiao Tong University, Shanghai 200240, China

[2] Stanford Synchrotron Radiation Lightsource, SLAC National Accelerator Laboratory, Menlo Park, CA 94025, USA

[3] Shanghai Motor Vehicle Inspection Certification & Technology Innovation Center Co., LTD, Shanghai 201805, China

[4] Instrument Analysis Center, Shanghai Jiao Tong University, Shanghai 200240, China

[5] Suzhou Institute of Nanotech and Nanobionics (SINANO), Chinese Academy of Sciences, Suzhou, Jiangsu 215123, China

[6] In-Situ Center for Physical Sciences, School of Chemistry and Chemical Engineering, Shanghai Jiao Tong University, Shanghai 200240, China

[7] National Synchrotron Light Source II, Brookhaven National Laboratory, Upton, NY 11973, USA

[8] Shanghai Jiao Tong University Sichuan Research Institute, Chengdu 610213, China

[#] These authors contributed equally to this work.

**Corresponding Author**: Yijin Liu (liuyijin@slac.stanford.edu) and Linsen Li (linsenli@sjtu.edu.cn)

**Abstract**

The electrification revolution in automobile industry and others demands annual production capacity of batteries at least on the order of $10^2$ gigawatts hours, which presents a twofold challenge to supply of key materials such as cobalt and nickel and to recycling when the batteries retire. Pyrometallurgical and hydrometallurgical recycling are currently used in industry but suffer from complexity, high costs, and secondary pollution. Here we report a direct-recycling method in molten salts (MSDR) that is environmentally benign and value-creating based on a techno-economic analysis using real-world data and price information. We also experimentally demonstrate the feasibility of MSDR by upcycling a low-nickel polycrystalline $LiNi_{0.5}Mn_{0.3}Co_{0.2}O_2$ (NMC) cathode material that is widely used in early-year electric vehicles into Ni-rich (Ni > 65%) single-crystal NMCs with increased energy-density (>10% increase) and outstanding electrochemical performance (>94% capacity retention after 500 cycles in pouch-type full cells). This work opens up new opportunities for closed-loop recycling of electric vehicle batteries and manufacturing of next-generation NMC cathode materials.

**Introduction**

Lithium-ion batteries (LIBs) play an important role in electrification revolution of the automobile industry and others and will continue to drive technology innovations.[1] More than 6 million electric vehicles (EVs) have been sold since 2010 and the sales are projected to rapidly increase in the coming years, which demands annual production capacity of LIBs at least on the order of 40 GWh yr$^{-1}$ or approximately 200, 000 metric tons of cathode materials annually.[2,3] Lithium nickel manganese cobalt ($LiNi_xMn_yCo_zO_2$, $x + y + z = 1$) and lithium nickel cobalt aluminum oxide are two widely used cathode materials for the EV batteries. The Ni content in NMCs has been gradually increasing (from 33% to ~90%) over the years due to the pursuit of a higher energy-density and a lower cost (replacing Co) for the EV application.[4] The growing demand for batteries has raised concerns on the sustainability due to limited cobalt and nickel resources on earth.[5,6] On the other hand, EV batteries commonly have a life-span ranging from 5–8 years. Therefore, a large quantity of used batteries are expected to be retired soon, especially those containing low-Ni (Ni ≤ 50%) NMCs deployed in the early-year EVs.[7] Given the significant economic and environmental impacts, there is a sense of urgency and growing momentum among governments, businesses, and consumers to establish closed-loop recycling so that the high-cost (accounting for ~40% of the total battery cost) and energy-intensive cathode materials can be re-integrated into the battery/EV manufacturing supply chain.[7-10]

Currently, EV batteries (using NMC or NCA) are recycled mainly through pyrometallurgical and hydrometallurgical approaches in industry.[7-9,11] Pyrometallurgy uses high temperature smelting (usually ≥1000 °C) to burn away organic materials such as polymer binder and separator, and produces alloys containing Ni, Co, and Cu through carbon reduction (**Figure 1** bottom). These high-value metals are further refined (often through hydrometallurgical processes) and recovered with high efficiency (>95%). Despite its simplicity and high productivity, pyrometallurgy is energy intensive and leads to high $CO_2$ emissions. Further, it is clearly not economically viable to synthesize new NMC materials starting from the recovered Ni and Co metals. By contrast, hydrometallurgical recycling operates at low temperature and employs multi-step chemical processes including leaching, extraction, and chemical precipitation to recover Ni, Co, Al, Cu, and Li from the used batteries (**Figure 1** top). Notably, high-purity Ni, Co, and Mn sulfates can be produced, which are the industry-standard precursors for production of NMC cathode materials through co-precipitation and sintering with lithium hydroxide or lithium carbonate. Even though hydrometallurgy enables a closed-loop recycling, it usually involves >10 major steps and generates a large amount of acid and base waste, which add to the cost and complexity.[7]

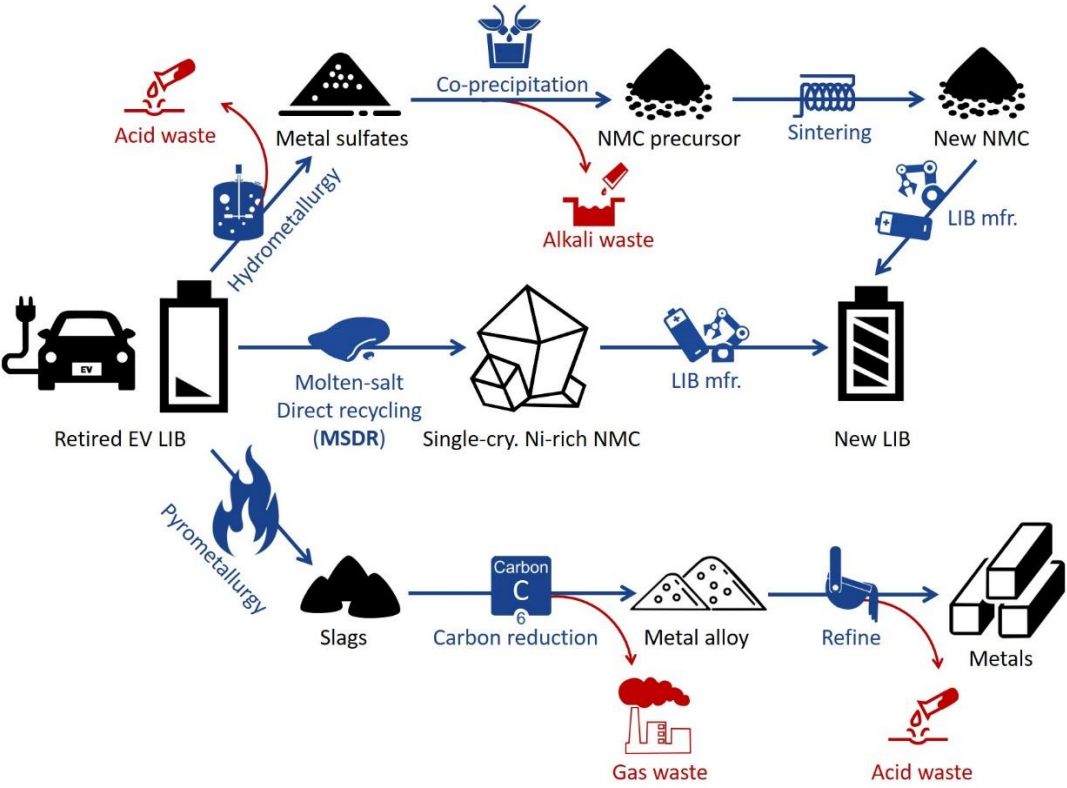

**Figure 1 | Comparison of different battery recycling technologies.** Hydrometallurgy and pyrometallurgy are currently operated in industry and suffer from complexity, high costs, and secondary pollution. Molten-salt direct-recycling (MSDR) is an environmentally benign process with fewer steps and for the first time provides the flexibility in tuning microstructure (polycrystalline to single-crystalline) and composition (low-Ni to Ni-rich) of the recycling-product to increase its value.

Direct recycling, which focuses on generating products that go directly back into new batteries without costly reprocessing and secondary pollution, has received increasing attention recently.[11-14] One major barrier with the existing direct-recycling technology is its inflexibility—what goes in comes out (i.e. same chemical composition and microstructure).[7] In practice, there will be little incentive for recycling if it does not create products with higher value to offset the processing cost. Furthermore, the inflexibility leads to a significant difficulty in addressing the mismatch between what need to be recycled first [low energy-density, low-Ni NMCs (Ni≤50%) that is widely used in early years and therefore would retire first, such as $LiNi_{1/3}Mn_{1/3}Co_{1/3}O_2$ and $LiNi_{0.5}Mn_{0.3}Co_{0.2}O_2$ (NMC532)] and what are preferred now and in the future [high energy-density Ni-rich (Ni ≥60%) NMCs, such as $LiNi_{0.6}Mn_{0.2}Co_{0.2}O_2$ and $LiNi_{0.83}Mn_{0.09}Co_{0.08}O_2$]. To address these challenges, here we report a direct-recycling technology through molten-salt chemistry that for the first time provides the flexibility in tuning the microstructure and the composition of the recycling-product to increase its value (MSDR; **Figure 1** middle). As a proof-of-concept, we use MSDR to upcycle degraded low-Ni polycrystalline NMC532 (Ni = 50%) retrieved from retired EV batteries (50 Ah capacity) into high-performance Ni-rich single-crystal NMCs, which receive significant interest recently from both academia[15] and industry[16]. This work opens up new opportunities for closed-loop recycling of EV batteries and manufacturing of next-generation NMC cathode materials.

**Results and discussion**

**Techno-economic analysis of the MSDR process**

We first analyze the costs for upcycling the degraded NMC532 from the used EV batteries into NMCs with different Ni contents (mole fraction of Ni = 50% to 90% in Ni+Co+Mn) through the MSDR process using best available information and compared the results with those from the hydrometallurgy route. We note that NMC532-graphite is the most widely used battery chemistry for EV batteries so far.[17] We did not carry out the calculation for the pyrometallurgy route because it is not a closed-loop recycling approach. We choose to use the prices from China (in Chinese Yuan ¥, 1 US dollars = 6.48 ¥), which holds the leading position in battery manufacturing capacity and has a large amount of information available. The numerical results of this techno-economic analysis are tabulated in **Supplementary Table 1**, including details of the calculations, including input parameters, key assumptions, and literature sources. According to our cost analysis, the MSDR process costs ~6000–10000 CNY ton$^{-1}$ less than the hydrometallurgical method in upcycling the degraded NMC532 to NMCs with different Ni contents (**Figure 2a**). The cost breakdowns and main factors contributing to the determination of each recycling product's production costs are shown in **Figure 2b.** Depending on the Ni content of the targeted final product, the cost fraction of the degraded NMC532 (red) and that of the added materials (blue) are negatively correlated. The added materials (blue) consist of two major components, Ni compounds and Li compounds, both of which are indispensable in the synthesis of Ni-rich cathode materials. The added amount of Ni compounds is clearly determined by the desired increment of Ni content in the final product. On the other hand, different Li compounds need to be incorporated in the synthesis protocol for different NMC products ($Li_2CO_3$ for Ni50-NMC and Ni60-NMC; LiOH for Ni70-NMC to Ni90-NMC), leading to a positive correlation of the material cost with the Ni content. Although the other cost components, e.g., labor cost, electricity and water utility, are less expensive comparing to the material costs, but they are quite significant to the cost reduction using our MSDR method and to the overall energy efficiency of the battery upcycling process. To further quantify this, we compose a pie chart to highlight the cost differences in the main factors between MSDR and hydrometallurgy, using Ni66-NMC (Ni = 66%) as an example (**Figure 2c**). The largest difference is in the utility of Electricity & Water (53.1%), which is followed by the cost of the added materials (18.1%) and cost of the degraded NMC532 from the retired batteries (16.4%). These cost reductions originate from the simplified MSDR synthesis procedure, which eliminates the usage of a number of chemicals and improves effectiveness of degraded NMC532 recycling. Single-variable sensitivity analyses for the production cost of Ni66-NMC is carried out to understand how different factors affect the total recycling costs through MSDR and hydrometallurgy, respectively (**Figure 2d**). The values shown in **Figure 2d** indicate the impacts of different factors under their respective market prices in the optimistic (denoted as Hydro positive and MSDR positive, shaded bars, low price) and pessimistic (denoted as Hydro negative and MSDR negative, solid bars, high price) scenarios. Overall, the MSDR method is cost-competitive over the hydrometallurgical method as indicated by the offset of the red-colored bars to the lower value in **Figure 2d**. There are three types of factors identified in this analysis. The first type is the price of LiOH·$H_2O$, which is quite significant and could potentially overwhelm MSDR's advantage in cost efficiency. The second type includes the price of NiSO4 and that of the degraded NMC532, in which the MSDR's cost advantage becomes more viable and is less sensitive to the market fluctuation. The third type includes the electricity utility and labor cost, in which the MSDR's advantage clearly prevail regardless of the market price variation. In spite of the uncertainties caused by the market fluctuation, from the statistical perspective, the MSDR method emerges as an economically attractive approach to recycling EV battery cathodes.

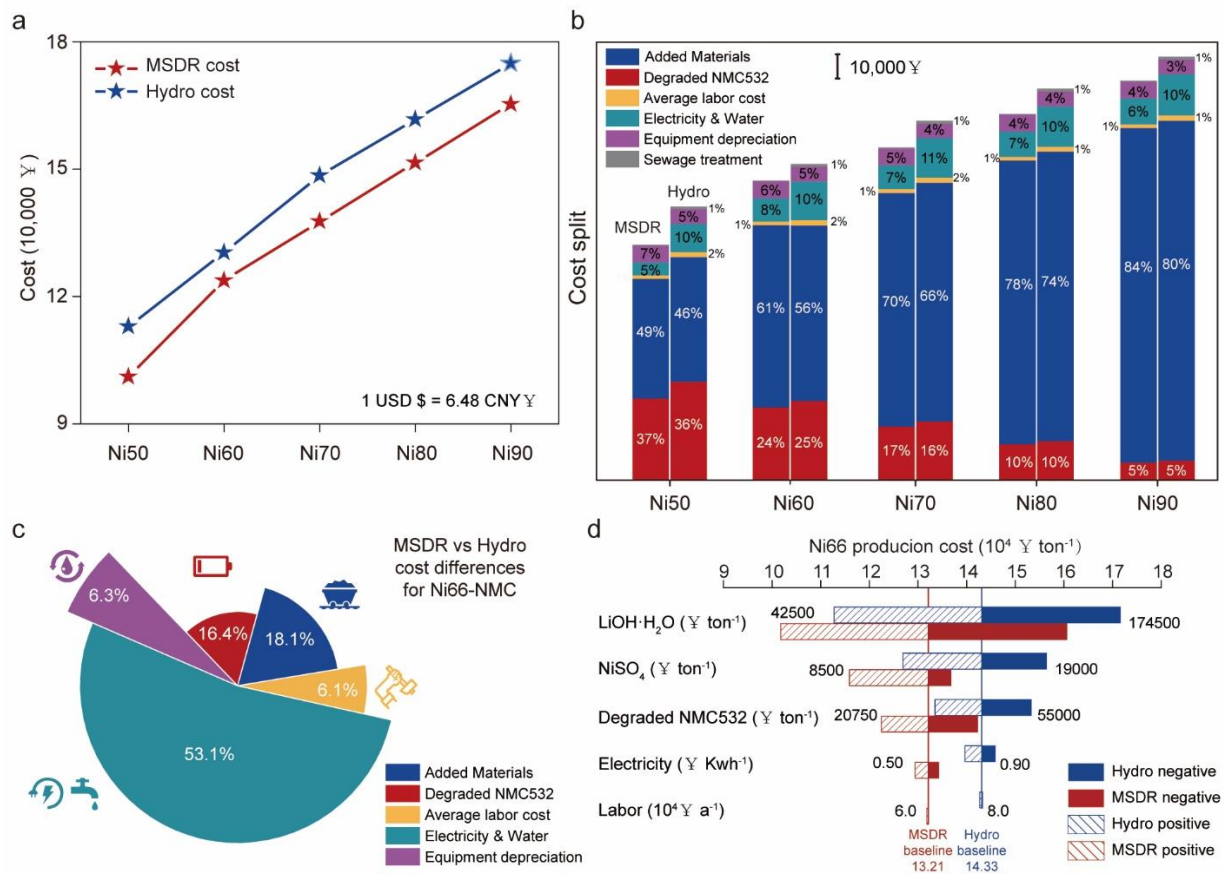

**Figure 2 | Techno-economical assessments of the MSDR and hydrometallurgical processes. a,** The costs of upcycling the degraded NMC532 production into NMCs with different Ni contents (mole fraction of Ni = 50% to 90%, Ni50 to Ni90) through MSDR and hydrometallurgy, respectively; **b,** Cost breakdowns for each of the targeted products; **c,** Pie chart showing the cost differences in main factors between MSDR and hydrometallurgical recycling. Upcycling of the degraded NMC532 into Ni66-NMC was used as an example; **d,** Single-variable sensitivity analysis for the production cost of Ni66-NMC through MSDR and hydrometallurgical recycling. The baseline parameters are calculated based on our techno-economic model shown in **Supplementary Table 1**. The values shown in this panel indicate the impacts of different factors under their respective market prices in the optimistic (denoted as Hydro positive and MSDR positive, shaded bars, low price) and pessimistic (denoted as Hydro negative and

MSDR negative, solid bars, high price) scenarios. Detailed cost breakdowns and sensitivity analysis are provided in the **Supplementary Table 2**.

**Diagnostics of the degraded NMC532 cathode from retired EV batteries**

To demonstrate of the feasibility of the MSDR method, we chose to recycle large-capacity (50 Ah) commercial batteries (**Figure 3a**, with NMC532 cathode and graphite anode) from the EVs that actually reached their end-of-life through real-world operation. The pouch cells used for this study were taken out of the battery pack and we conduct one cycle of charge and discharge at a current of 5 A to determine its remaining capacity. The cell retained 73% of its initial nominal capacity (**Figure 3b**). After disassembling the cell and isolating the cathode powder (see procedures in **Methods**), we diagnose the chemical and structural degradations of the NMC532 cathode. Induced coupled plasma-optical emission spectroscopy (ICP-OES) analysis reveals a molar ratio of 0.77:1 between Li and Ni+Mn+Co in the degraded NMC532 (**Figure 3b**). Cross-sectional scanning electron microscopy (SEM, **Figure 3c** and **3d**) and X-ray micro-computed-tomography (X-ray micro-CT, **Supplementary Fig. 1**) clearly reveal severe intergranular fractures in the polycrystalline NMC532 particles. These microcracks are closely related to capacity decay and power loss of the NMC cathodes, which has been discussed extensively in literature.[18-20] X-ray diffraction (XRD) pattern suggests the loss of the layered structure and the phase transitions in the degraded NMC532, as indicated by the emergence of (006) and (012) diffraction peaks and the collective peak shifts (**Figure 3e**).[21-23] These observations are further confirmed by high-angle annular dark-field imaging using a scanning transmission electron microscope (HAADF-STEM, **Figure 3f**), which clearly shows the rocksalt-like regions near the particle surface (see the fast-Fourier transform diffractogram in **Figure 3g**). Layered structure can still be observed in the bulk (**Figure 3f** and **3h**). At the sub-surface region, we also detected transition metal (TM) occupation in the Li layer (annotated by the white arrows in **Figure 3f**), which is known to impede $Li^+$ transport.[24] From the above results, we conclude that the NMC532 cathode from the retired EV batteries is lithium deficient, structurally degraded, and morphologically damaged.

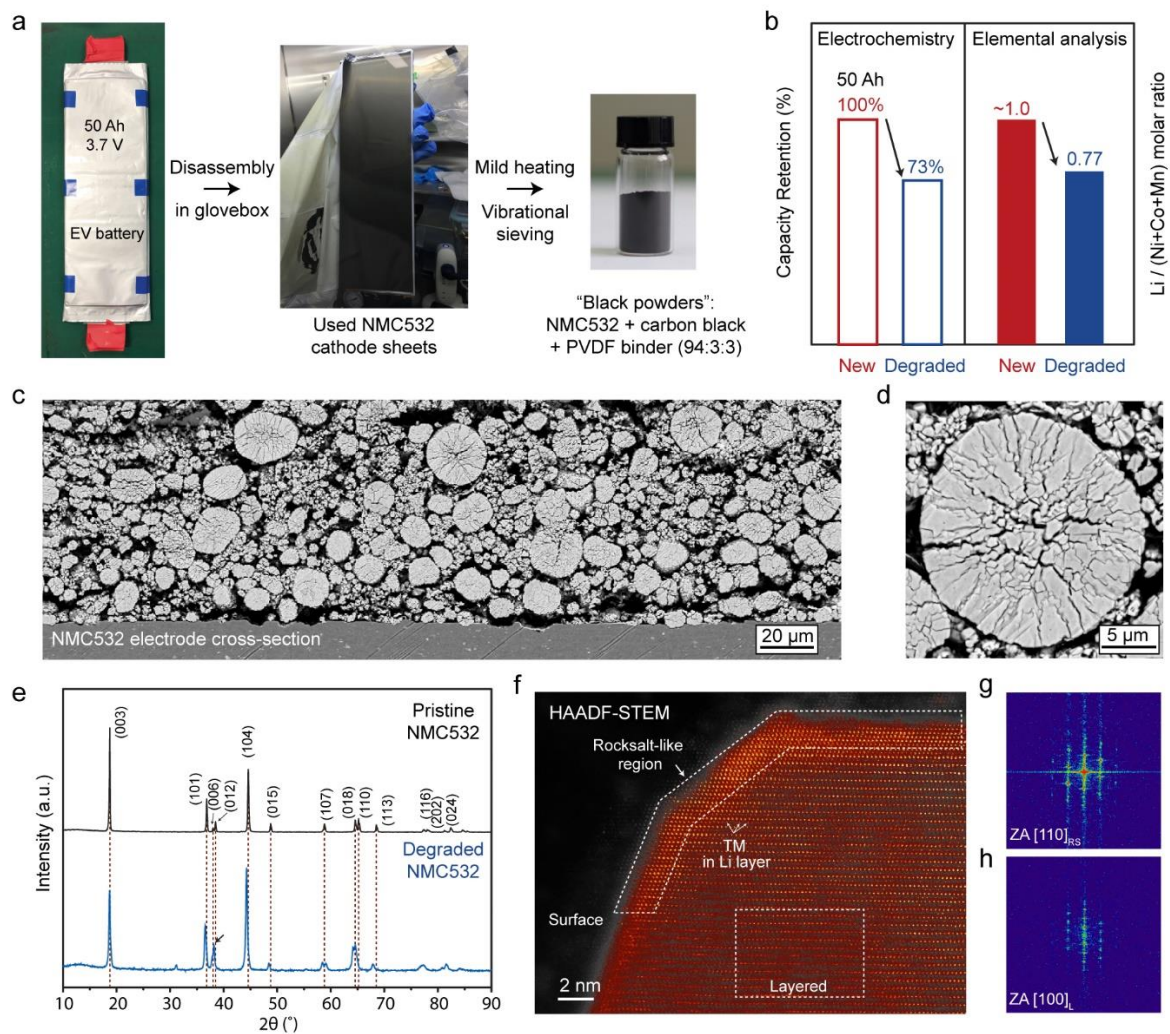

**Figure 3 | Diagnostics of the degraded NMC532 cathode from retired EV batteries a,** a 50 Ah pouch cell retrieved from battery packs of retired electric vehicles and its disassembly processes; **b,** Electrochemical capacity quantification of the 50 Ah pouch cell (left panel) and the lithium loss in the degraded $LiNi_{0.5}Mn_{0.3}Co_{0.2}O_2$ (NMC532) cathode material measured by ICP-OES (right panel). **c** and **d** are cross-sectional SEM images of the double-side coated cathode showing severe intergranular fracturing in the degraded polycrystalline NMC532 particles, which leads to significant increase in surface area and is known to cause battery-performance; **e,** XRD patterns of the pristine NMC532 material powder and the degraded one from the retired pouch cell; **f,** HAADF-STEM image of the degraded NMC532 particle, showing the rocksalt-like regions near the surface. The layered structure can still be observed in the bulk region; Panel **g** and **h** are FFT diffractograms from the rocksalt-like and the layered regions, respectively.

**Ni-rich single-crystal NMCs prepared through MSDR**

Although re-lithiation methods in low-temperature eutectic molten salts[13] or under hydrothermal conditions[25] for direct recycling have been previously reported to restore the stoichiometric composition and crystal structure of the degraded NMC532 materials, they cannot heal the microcracks. Neither do they provide the flexibility in readjusting the elemental composition or microstructure of the recycling products to increase the economic feasibility of the recycling process. We notice that many transition metal oxides (such as NiO) are soluble in molten salts[26,27] and take advantage of this phenomenon to develop the herein presented MSDR method for upcycling low-Ni polycrystalline NMC into Ni-rich single-crystal NMC. In the MSDR process, the cathode powders from the used battery, which contains degraded NMC532 particles, carbon black, and polymer binder (NMC cathode material typically ≥94 wt%), are used as "materials feedstock". They are first mixed with Ni-rich materials such as $Ni_{0.83}Mn_{0.09}Co_{0.08}(OH)_2$ or $Ni(OH)_2$ (SEM images shown in **Supplementary Fig. 2**) and then heated at a high temperature in $LiOH$-$Li_2SO_4$ molten-salt mixtures (see details in **Methods**). The small amount of carbon black and polymer binder in the NMC532 cathode (<6 wt%) are burned away during the thermal treatment. A time-resolved *in situ* XRD experiment is carried out to track the reaction between equal moles of the degraded NMC532 (Ni = 50%) and $Ni_{0.83}Mn_{0.09}Co_{0.08}(OH)_2$ (Ni = 83%) in molten $LiOH$-$Li_2SO_4$ (**Figure 4a**). The targeted composition is $Li_{1.0}Ni_{0.665}Mn_{0.195}Co_{0.140}O_2$ (denoted by Ni66-NMC). At the beginning, peaks associated with the degraded NMC532, $Ni_{0.83}Mn_{0.09}Co_{0.08}(OH)_2$, LiOH, and $Li_2SO_4$ could all be observed. As the temperature increase, $Ni_{0.83}Mn_{0.09}Co_{0.08}(OH)_2$ peaks disappear at 200 °C, likely due to the loss of $H_2O$ and long-range structural ordering. The LiOH and $Li_2SO_4$ peaks exist until ~300 °C. As the temperature and holding-time further increase, the (003) and (104) peaks associated with the layered structure gradually increase in intensity. The initially merged (006) and (012) peaks in the degraded NMC532 become separated, which also indicates the recovery of the layered structure. The overall chemical composition of the NMC product after the MSDR process is determined to be $Li_{1.02}Ni_{0.66}Mn_{0.19}Co_{0.13}O_2$ by ICP-OES, which is quite close to the targeted composition ($Li_{1.0}Ni_{0.665}Mn_{0.195}Co_{0.140}O_2$, Ni66-NMC). Compared with the degraded NMC532, the recycling product become fully lithiated and Ni-rich. The structural and chemical transformations of the NMC during the MSDR process are schematically summarized in **Figure 4b**. Ni66-NMC powders are obtained after cooling and water-washing to remove the water-soluble Li salts, which can be reused after water evaporation (**Supplementary Fig. 3**).

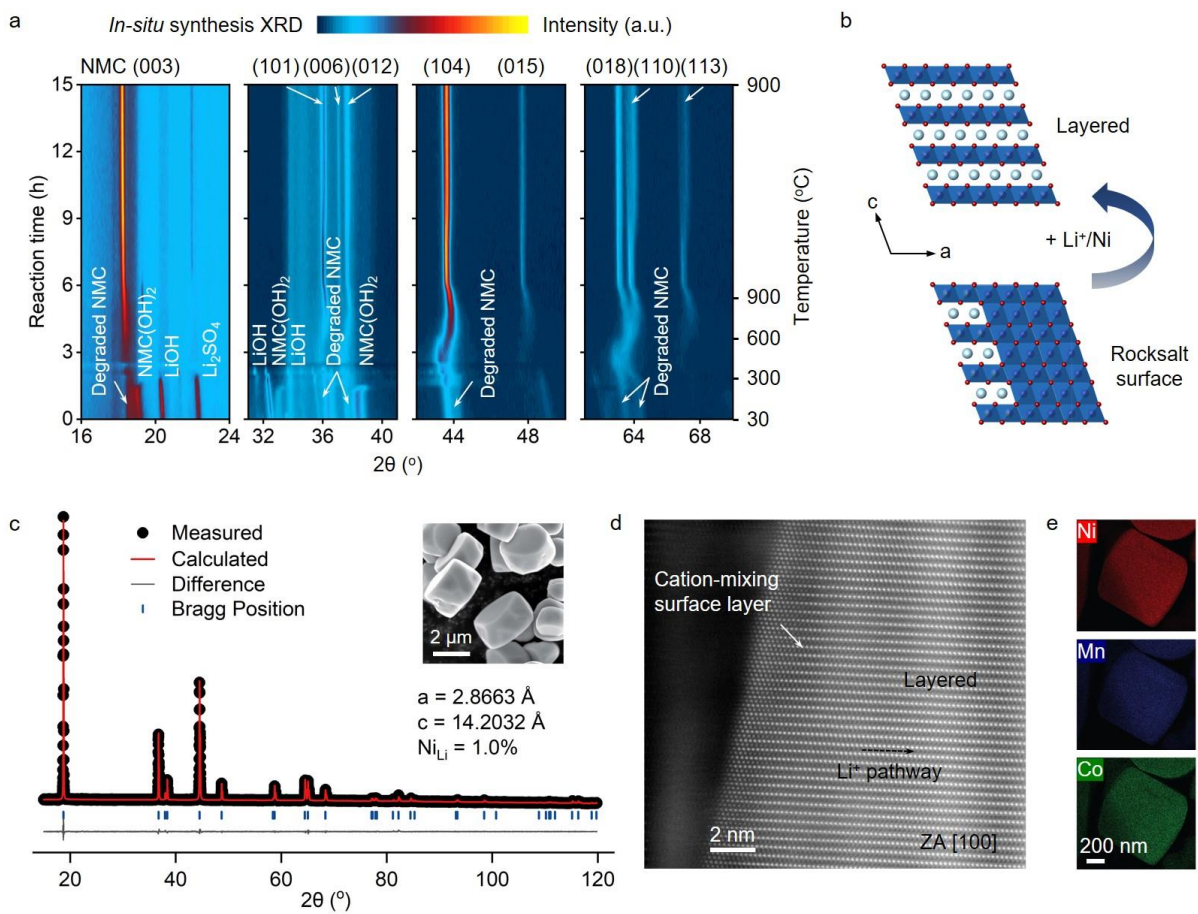

**Figure 4 | Single-crystal Ni66-NMC material prepared through MSDR. a,** Two-dimensional contour plot of the *in situ* XRD patterns recorded during the conversion of the degraded polycrystalline NMC532 into a single-crystal Ni-rich NMC (Ni = 66.5%) through MSDR. The degraded NMC532 was mixed with a Ni-rich compound, $Ni_{0.83}Mn_{0.09}Co_{0.08}(OH)_2$ [denoted by $NMC(OH)_2$ in the graph] and heated at 900 °C in the $LiOH$-$Li_2SO_4$ molten salts in air; **b,** Schematic illustration show that the layered structure is restored after the MSDR process based on the *in situ* XRD results; **c,** High-resolution powder XRD of the Ni66-NMC material and the Rietveld refinement results. The Ni66-NMC is phase-pure and has a highly-ordered hexagonal $\alpha$–$NaFeO_2$ structure with $R\bar{3}m$ symmetry. The $Ni_{Li}$ cation mixing is estimated at ~1.0 %. Inset

shows the SEM image of the single-crystal Ni66-NMC particles; **d,** HAADF-STEM image collected from the surface of a Ni66-NMC particle along its [100] zone-axis, showing a highly ordered layered structure in the bulk and a surface cation-mixing layer of several nm in thickness. **e,** STEM-EDS elemental mapping of a Ni66-NMC particle showing fairly uniform distribution of Ni, Mn, and Co in the thin-slice sample.

High-resolution XRD pattern collected from the Ni66-NMC material reveals a single phase that exhibits a highly-ordered hexagonal $\alpha$–NaFeO$_2$ structure with $R\bar{3}m$ symmetry (**Figure 4c**). Rietveld refinement is performed on the full XRD data, suggesting an estimated Ni$_{Li}$ cation mixing of ~1.0 %, which is relatively low compared with the typical values for NMC cathode materials (2–4%).[28] The morphology of the Ni66-NMC is examined by SEM (inset in **Figure 4c**), featuring well-faceted particles at about 2–4 μm. Some particles are single-grained while others consist of several agglomerated grains. Such morphology is consistent with that of the single-crystal NMCs reported in literature.[29] The difference in the morphology between the degraded polycrystalline NMC532 particles (**Figure 3c** and **3d**) and the synthesized single-crystal Ni66-NMC particles (inset in **Figure 4c**) suggests that the MSDR process likely follows a dissolution-reaction-crystallization mechanism. **Figure 4d** shows a representative STEM image collected along [100] zone-axis of a Ni66-NMC particle, which has a highly ordered layered structure in the bulk and a surface cation-mixing layer (several nm in thickness). Energy-dispersive X-ray spectroscopy (EDS) elemental mapping using the STEM reveals fairly homogeneous distributions of Ni, Mn, and Co with an as-designed stoichiometric ratio in the thin-slice (~50 nm in thickness) sample used in the STEM characterization.

It is critical for the NMC particles to have homogeneous elemental compositions so that they can show a high degree of consistency in the electrochemical tests.[23] Therefore, nano-resolution X-ray fluorescence (XRF) tomography is further performed using a synchrotron hard X-ray nanoprobe instrument to quantitatively analyze the particle-level compositional variation of the Ni66-NMC material synthesized through our recycling protocol. XRF can achieve higher precision and sensitivity than STEM-EDS (~10–100 ppm for XRF versus ~1 atomic% for STEM-EDS).[30] **Figure 5a** shows the reconstructed 3D tomogram of a randomly selected Ni66-NMC particle. This particle exhibits a well-formed shape and is rather uniformly crystallized as indicated by the uniform particle density, which is demonstrated with the virtually sliced views shown in **Figure 5a**. Some localized regions, however, show a mild density variation, which motivates an in-depth investigation of the compositional distribution over the corresponding regions. **Figure 5b** shows Ni, Co, and Mn elements' relative concentration distribution. While the Ni distribution is fairly homogeneous, the Co and Mn elements demonstrate a "Core-Shell" separation. Co is enriched in the shell (see arrows in **Figure 5b**) and its concentration becomes lower as it approaches the core region. The spatial distribution of Mn seems to be inversely correlated with Co. The averaged composition over this entire particle is quantified in **Figure 5c**, in good agreement with the bulk-averaged value. This Core-Shell compositional variation is further supported by our depth-dependent cation concentration plot shown in **Figure 5d**, which is also echoed by the Co's shoulder peak annotated in **Figure 5c**. It has been reported that, the presence of Co could suppress the cation mixing and, on the contrary, Mn could promote this undesired phenomenon.[23] The Co-rich surface could, therefore, suppress the moisture-induced detrimental surface phase transition. In addition, during the recycling process, some single-crystal particles inevitably agglomerate together to form clusters. To understand the effect of particle agglomeration, we further analyze the compositional distribution over a three-particle cluster (**Supplementary Figure 4**). It is interesting that the "Mn-rich-Core/Co-rich-Shell" feature clearly persists in the imaged cluster (**Supplementary Figure 4**), although some minor particle-to-particle variations can be observed.

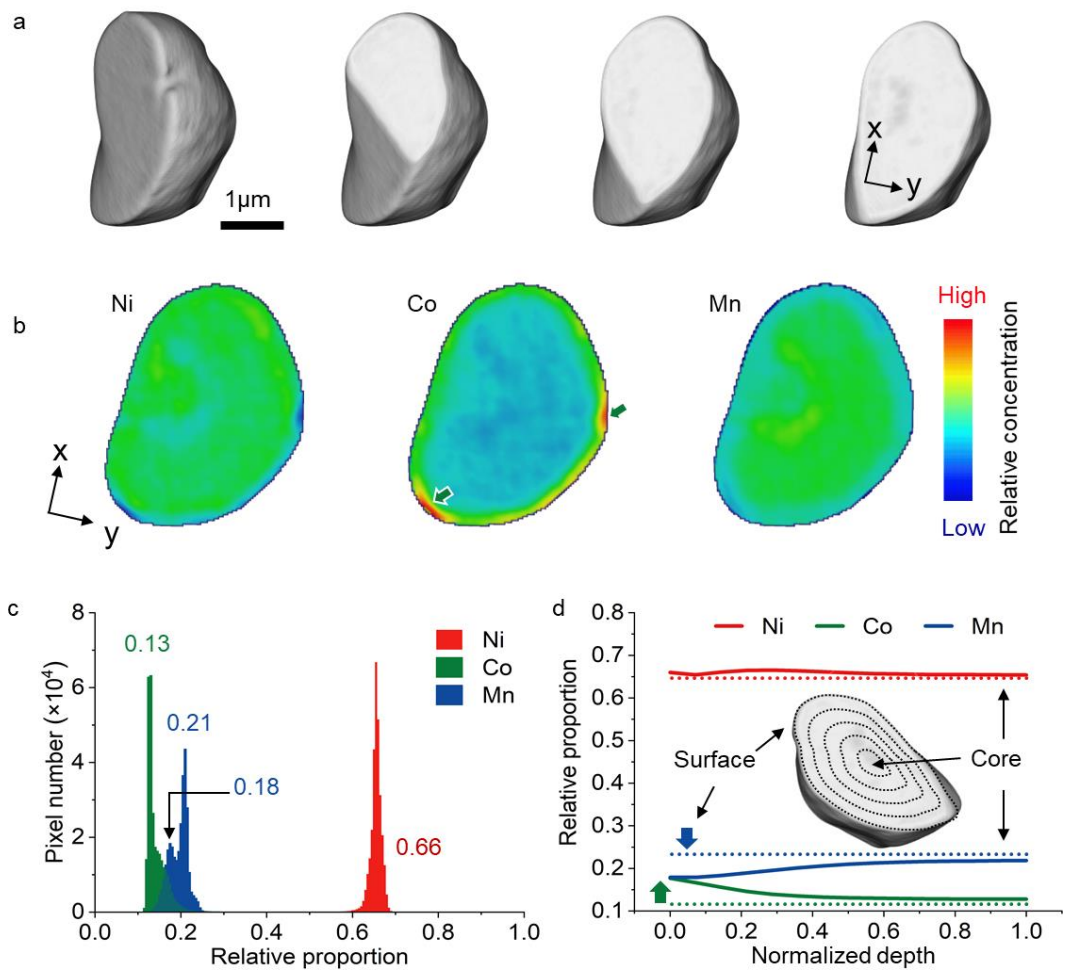

**Figure 5 | The elemental analysis of the MSDR-synthesized single-crystal Ni66-NMC particle. a,** The 3D morphology of an randomly selected Ni66-NMC particle with its internal structure visualized through virtual slicing; **b,** Relative concentrations of Ni, Mn, and Co over a virtual 2D slice along *xy*-plane revealed by X-ray fluorescence (XRF) tomography; **c,** The histograms of the elemental concentrations over the entire Ni66-NMC particle; **d,** The depth-dependent profiles of the transition metal distribution over the imaged Ni66-NMC particle.

The demonstration of the targeted Ni content (~66.5% in this case) from the electrode-level (ICP-OES data) to the individual-particle-level (EDS data and XRF tomography data) provides strong evidences for the dissolution-reaction-crystallization mechanism proposed for the MSDR process. The overall reaction equation can be written as following.

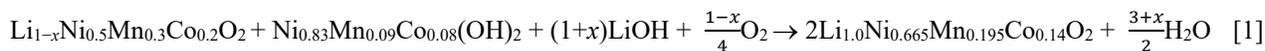

$$Li_{1-x}Ni_{0.5}Mn_{0.3}Co_{0.2}O_2 + Ni_{0.83}Mn_{0.09}Co_{0.08}(OH)_2 + (1+x)LiOH + \frac{1-x}{4}O_2 \rightarrow 2Li_{1.0}Ni_{0.665}Mn_{0.195}Co_{0.14}O_2 + \frac{3+x}{2}H_2O \quad [1]$$

The molten LiOH-Li$_2$SO$_4$ functions as the reaction medium for ion transport and crystal growth. All the Ni, Co, and Mn and the remaining Li from the degraded NMC532 are used as "building-blocks" in the MSDR process, which indicates a high atom-economy (~100%). The actual yield is calculated to be >95% (from >10 separate batches). The loss is primarily due to the small amount of Ni66-NMC powders stuck to the bottom of the reaction crucible. To further demonstrate compositional tunability through MSDR, the degraded NMC532 (Ni = 50%) was mixed with Ni(OH)$_2$ (Ni = 100%) at a molar ratio of 2:3 to synthesis a Ni-rich NMC with a targeted Ni content of 80% (LiNi$_{0.8}$Mn$_{0.12}$Co$_{0.08}$O$_2$, denoted by Ni80-NMC). Single-crystal Ni80-NMC particles are indeed produced (**Supplementary Fig. 5**). The actual composition is determined to be Li$_{1.010}$Ni$_{0.796}$Mn$_{0.123}$Co$_{0.081}$O$_2$ by ICP-OES, which is close to the targeted composition (LiNi$_{0.8}$Mn$_{0.12}$Co$_{0.08}$O$_2$). The overall reaction equation for this process can be written as following.

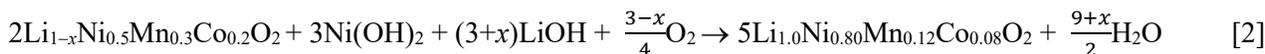

$$2Li_{1-x}Ni_{0.5}Mn_{0.3}Co_{0.2}O_2 + 3Ni(OH)_2 + (3+x)LiOH + \frac{3-x}{4}O_2 \rightarrow 5Li_{1.0}Ni_{0.80}Mn_{0.12}Co_{0.08}O_2 + \frac{9+x}{2}H_2O \quad [2]$$

**Electrochemical performance of the recycling single-crystal NMCs**

The electrochemical performance of the single-crystal NMCs prepared through MSDR was first investigated in half cells using Li foil as the counter electrode. **Figure 6a** shows the voltage profile of the degraded NMC532 cathode material from the retired EV batteries in comparison with that of a pristine one. The first-cycle voltage profiles of the Ni66-NMC and the Ni80-NMC cathodes prepared by upcycling of the degraded NMC532 are shown in **Figure 6b** and **6c**, respectively. With increasing Ni content, the first-cycle discharge capacity of the NMC cathodes increases in the order of the pristine NMC532 (161 mAh g$^{-1}$), Ni66-NMC (179 mAh g$^{-1}$) and Ni80-NMC (200 mAh g$^{-1}$) and so does the energy density. Consistent with previous reports on the single-crystal NMCs,[16,28,31] our single-crystal Ni66-NMC cathodes also demonstrate outstanding cycling stability in half-cell tests at 1 C charge and discharge. It retains ~95% of its initial capacity after 200 cycles at 30 °C (**Figure 6d**) in the voltage range of 4.3–2.8 V, a typical voltage window for testing NMC cathodes. When a more aggressive cycling condition is used (a higher cut-off voltage of 4.5 V and a higher temperature of 60 °C), the Ni66-NMC cathode can still retain ~80% of its initial capacities after 200 cycles (**Figure 6e**). The Ni80-NMC cathode also shows a decent cycling performance by retaining 85% of its initial capacity after 200 cycles (**Figure 6f**). In order to further evaluate the long-term cycling stability of single-crystal Ni66-NMC, pouch-type full cells are assembled using artificial graphite as the anode and cycled in the voltage range of 4.2–2.7 V at 0.5 C after three formation cycles at 0.1 C (**Figure 6g**). The Ni66-NMC|graphite full batteries retain >94% of its initial capacity after 500 cycles (results of three pouch cells are shown in **Figure 6h**). If we assume the same capacity decay rate, the cycle life is projected to approaching ~2000 cycles (till 80% capacity retention), which is comparable to the state-of-the-art performance for the Ni-rich NMC cathodes[16]. The Ni66-NMC cathode also shows excellent rate capability and retains a discharge capacity of 136 mAh g$^{-1}$ at 20 C (**Supplementary Figure 6**). These electrochemical results strongly support that the MSDR process can produce high-performance and thus high-value single-crystal Ni-rich NMC cathode materials from the degraded low-Ni NMC532.

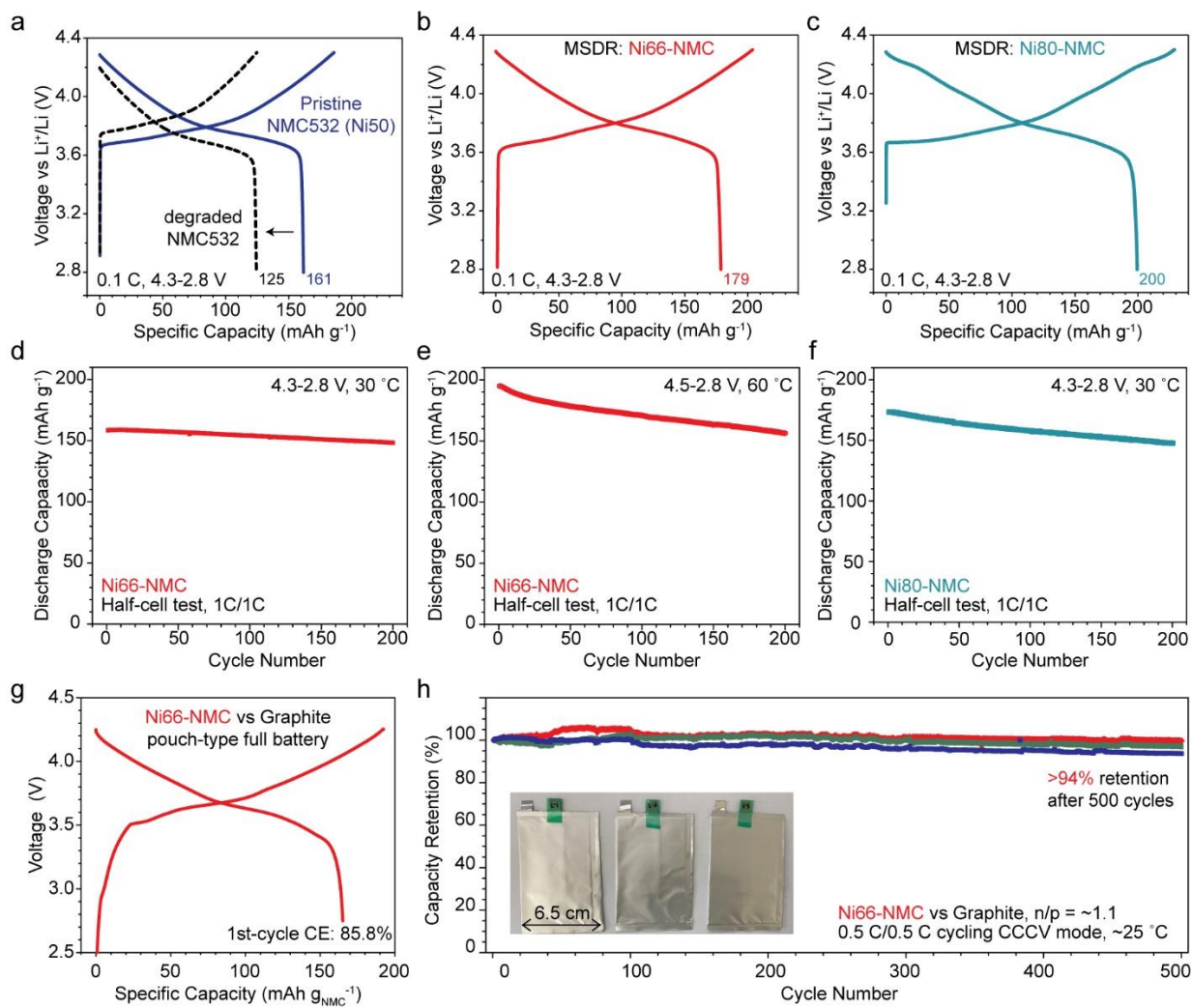

**Figure 6 | Electrochemical performance of the MSDR-recycled single-crystal Ni-rich NMCs a,** Voltage profiles of a pristine NMC532 cathode and the one retrieved from the retired EV batteries tested at 0.1 C and 30 °C; **b,** Voltage curves of the Ni66-NMC cathode in the first cycle tested at 0.1 C and 30 °C; **c,** Voltage curves of the Ni80-NMC cathode in the first cycle tested at 0.1 C and 30 °C; **d,** Cycling performance of the Ni66-NMC under normal (4.3–2.8 V, at 30 °C) and **(e)** aggressive testing conditions (4.5–2.8 V, at 60 °C) at 1 C; **f,** Cycling performance of the Ni80-NMC cathode; **g,** Voltage profiles of a pouch-type Ni66-NMC|graphite full battery in the first cycle at 0.1 C; **h,** Long-term cycling performance of three pouch-type full cells cycled at 0.5 C in the voltage of 4.2–2.7 V, showing >94% capacity retention after 500 cycles.

**Conclusion**

The rapid growth of the global battery market requires closed-loop recycling technologies, which are critical to building a sustainable and clean energy future. An economically viable recycling process needs to be value creating. To this end we have developed an upcycling technology that converts low-Ni polycrystalline NMC cathodes from retired EV batteries into Ni-rich single-crystal cathodes through molten-salt chemistry. This method is more efficient and environmentally benign compared to the traditional hydrometallurgical and pyrometallurgical approaches and is cost-competitive according to our techno-economic analysis based on real-world data and price information. Our MSDR recycling protocol has demonstrated precise tunability for the final products' elemental compositions from the electrode level down to the single grains level. The synthesized single-crystal Ni-rich cathodes exhibit an outstanding electrochemical performance in practically relevant full battery tests. Our experimental results and techno-economic analysis demonstrate the great potential of MSDR in closed-loop recycling of the EV batteries and in reshaping the manufacturing of next-generation NMC cathode materials.

**Methods.**

*EV battery test and disassembly:*

50 Ah pouch-type EV batteries were taken out from the battery packs of the retired EVs. The pouch cell used in this work was first charged and discharged at a current of 5 A for one cycle. The used batteries were obtained typically had a capacity retention of ~65–85% (relative to the nominal capacity, 50 Ah). The tabs were then insulated by tapes to avoid unintentional short-circuits. The pouch cell was transferred into an argon-filled glove box ($O_2/H_2O$ <0.1 ppm). The pouch made of aluminum laminated film was cut open and the *Z*-folded stack of cathode/separator/anode was taken out. The double-side coated cathode sheets (made of NMC532, carbon black, and polymer binder) were separated from the stack and rinsed with fresh dimethyl carbonate to remove the electrolyte. The cathode sheets were thermally treated at 400 °C for 1 h to decompose the polymer binder so that the cathode powder could be separated from the aluminum foil by vibrational sieving. The cathode powder was collected and stored in the glovebox before any further tests.

*MSDR of the degraded NMC532:*

To upcycle the degraded NMC532 ($Li_{1-x}Ni_{0.5}Mn_{0.3}Co_{0.2}O_2$) into Ni66-NMC, 0.2 mole of the degraded NMC532 cathode powder (~20 g, a mixture of NMC532, carbon black, and polymer binder) was mixed with 0.2 mole of $Ni_{0.83}Mn_{0.09}Co_{0.08}(OH)_2$ and 0.5 mole of LiOH and 0.1 mole of $Li_2SO_4$ and transferred into a crucible. LiOH was in excess [excess% = (1.5− $x$)/(1 + $x$)] to facilitate complete relithiation, which is a self-saturated process accord to literature[13]. The crucible used for reaction was loosely covered by a lid to allow air intake during the synthesis but avoid excess evaporation of Li salts. The material mixture was heated to 900 °C at a ramping rate of ~10 °C min$^{-1}$ and held at 900 °C for 5 h and then at 860 °C for 15 h in air before slow cooling at a rate of 2 °C min$^{-1}$ to 300 °C. After that, the mixture was allowed to cool down naturally. The obtained powders were recovered from the crucible, ground in an agate mortar, and washed using deionized water to separate the material from the water-soluble Li-salts, which could be recycled after water evaporation and reused. The Ni66-NMC powder was collected by centrifugation and dried at 70 °C in air before thermally treated at 700 °C for 4 hours in air. The powder was ground again to pass through a 400-mesh sieve and then stored in a humidity-controlled storage chamber before further tests. To upcycle the degraded NMC532 into Ni80-NMC via MSDR, 0.2 mole of the degraded NMC532 cathode powder (~20 g) was mixed with 0.3 mole of $Ni(OH)_2$ and 0.625 mole of LiOH and 0.125 mole of $Li_2SO_4$ and transferred into a crucible. The experimental procedures were the same as those for the Ni66-NMC except the sintering temperature was lowered to 780 °C.

*Structural Characterizations:*

XRD was performed using an X-ray diffractometer (Bruker D8, Cu K$\alpha$ radiation). SEM was carried out on a Phenom Pro microscope. SEM-EDS was carried out on a on a Zeiss Crossbeam 540 microscope. X-ray MicroCT was performed on Zeiss Xradia 520 Versa. ICP-OES was carried out on an iCAP™ 7600 ICP-OES Analyzer (Thermo Fisher). Ion-milling was performed using an ion beam slope cutter (Leica

EM TIC 3X). STEM specimen preparation was conducted by FIB-SEM on a Zeiss Crossbeam 540. The FIB-prepared samples were investigated by a Cs-corrected JEOL JEM-ARM200F operated at 200 kV. The nano-resolution fluorescence mapping was performed using the hard X-ray nanoprobe (HXN) at beamline 3-ID of the National Synchrotron Light Source II at Brookhaven National Laboratory. For the nano-fluorescence mapping, the coherent X-ray of 12 keV was focused to a spot size of 12 nm and >50 projections were collected. The sample's XRF signals (Ni, Mn, and Co) were converted to numbers of atoms using the measurements from XRF calibration standards (Micromatter 41149, 41157 and 41159). *In situ* XRD was performed using an X-ray diffractometer (Bruker D8, Cu K$\alpha$ radiation) equipped with a ceramic heating stage. The materials were mixed and pressed into a pellet, which was loaded onto the ceramic stage and heated from 30 °C to 900 °C at a ramping rate of ~2.5 °C min$^{-1}$. The XRD patterns were collected continuously and each scan took ~10.5 min.

*Electrochemical measurements:*

The NMC powders (NMC532, Ni66, and Ni80-NMC) were mixed with PVDF and super-P carbon black with a weight ratio of 90:5:5 in NMP using a Thinky Mixer (ARE-310). The slurries were coated onto aluminum foils using a film applicator. The electrodes were first dried in a convection oven at 80 °C for 2 hours and then dried at 120 °C under vacuum overnight. A typical loading of the electrodes for half-cell tests is about ~2–3 mg cm$^{-2}$. Half-cells were assemble using Li foil (0.45 mm in thickness) as the anode, a polyethylene separator, and 1.2 M LiPF$_6$ dissolved in EC-EMC (3/7 w/w) + 2 wt% VC as the electrolyte in an argon-filled glove-box O$_2$/H$_2$O <0.1 ppm). Electrochemical measurements were performed on coin cells inside constant-temperature chambers set at 30 °C using battery cyclers. The half-cells were cycled at 0.1 C for three cycles before cycling at 1 C (1 C = 160 mA g$^{-1}$ for NMC532, 180 mA g$^{-1}$ for Ni66-NMC, and 190 mA g$^{-1}$ for Ni80-NMC). In the rate capability test, the half-cell was cycled at 0.1 C for one cycle and then at 0.2, 0.5, 1, 2, 3, 5, 10, and 20 C each for five cycles. For the full-cell tests in the pouch cells, commercial artificial graphite was used as the anode material. The cathode loading level was ~8 mg cm$^{-2}$; the anode loading level was around ~4 mg cm$^{-2}$. The N/P ratio was adjusted to ~1.1. The electrodes for making full-cells were calendared using a roller press. The full cells (~30 mAh-capacity single-layer pouch cells) were cycled at 0.1 C for three cycles before cycling at 0.5 C at ~25 °C. The pouch cells were sandwiched by a pair of metal plates during testing.

**Acknowledgements.** This work is also supported by the Natural Science Foundation of China (22008154 to L.S.L.), the Sichuan Science and Technology Program (2021JDRC0015), and partially by Sinopec (420038–1, to L.S.L) and Hitachi Chemical (Shanghai) Co., Ltd. This research used the Hard X-ray Nanoprobe (HXN) Beamline at 3-ID of the National Synchrotron Light Source II, a U.S. Department of Energy (DOE) Office of Science User Facility operated for the DOE Office of Science by Brookhaven National Laboratory under Contract No. DESC0012704. Stanford Synchrotron Radiation Lightsource, SLAC National Accelerator Laboratory, is supported by the U.S. Department of Energy (DOE), Office of Science, Office of Basic Energy Sciences under Contract No. DE-AC02-76SF00515.

**Author Contribution.** L.S. L., G.N.Q., and Y.S.H. conceived the research. G.N.Q. performed the studies with help from other authors. Y.W. prepared the pouch cells. Y.S.H, X.Y.X., and H.Y.C. (EV battery test and disassembly), S.J.X., Y.B.S. and L.W.C. (XRD), Y.H.Z., J.Z.L., X.H. and Y.L. (X-ray imaging), and Z.Y.L. (techno-economic analysis). L.S.L., Y.L. and G.N.Q. wrote the manuscript with inputs from all other authors.

**Competing interests:** The authors declare no competing financial interest.

**Materials & Correspondence.** Correspondence and materials request should be addressed to Y.J.L. (liuyijin@slac.stanford.edu) and L.S.L (linsenli@sjtu.edu.cn).

**Reference**
1. Battery revolution to evolution. *Nat. Energy* **4**, 893-893 (2019).
2. R. E. Ciez & J. F. Whitacre. Examining different recycling processes for lithium-ion batteries. *Nat. Sustain.* **2**, 148-156 (2019).
3. G. Harper, R. Sommerville, E. Kendrick, L. Driscoll, P. Slater, R. Stolkin, A. Walton, P. Christensen, O. Heidrich, S. Lambert, A. Abbott, K. Ryder, L. Gaines & P. Anderson. Recycling lithium-ion batteries from electric vehicles. *Nature* **575**, 75-86 (2019).
4. Y. Ding, Z. P. Cano, A. Yu, J. Lu & Z. Chen. Automotive Li-Ion Batteries: Current Status and Future Perspectives. *Electrochem. Energy Rev.* **2**, 1-28 (2019).
5. E. A. Olivetti, G. Ceder, G. G. Gaustad & X. Fu. Lithium-Ion Battery Supply Chain Considerations:

Analysis of Potential Bottlenecks in Critical Metals. *Joule* **1**, 229-243 (2017).
6. C. Banza Lubaba Nkulu, L. Casas, V. Haufroid, T. De Putter, N. D. Saenen, T. Kayembe-Kitenge, P. Musa Obadia, D. Kyanika Wa Mukoma, J.-M. Lunda Ilunga, T. S. Nawrot, O. Luboya Numbi, E. Smolders & B. Nemery. Sustainability of artisanal mining of cobalt in DR Congo. *Nat. Sustain.* **1**, 495-504 (2018).
7. M. Chen, X. Ma, B. Chen, R. Arsenault, P. Karlson, N. Simon & Y. Wang. Recycling End-of-Life Electric Vehicle Lithium-Ion Batteries. *Joule* **3**, 2622-2646 (2019).
8. J. Heelan, E. Gratz, Z. Zheng, Q. Wang, M. Chen, D. Apelian & Y. Wang. Current and Prospective Li-Ion Battery Recycling and Recovery Processes. *JOM* **68**, 2632-2638 (2016).
9. X. Zhang, L. Li, E. Fan, Q. Xue, Y. Bian, F. Wu & R. Chen. Toward sustainable and systematic recycling of spent rechargeable batteries. *Chem. Soc. Rev.* **47**, 7239-7302 (2018).
10. E. Fan, L. Li, Z. Wang, J. Lin, Y. Huang, Y. Yao, R. Chen & F. Wu. Sustainable Recycling Technology for Li-Ion Batteries and Beyond: Challenges and Future Prospects. *Chem. Rev.* **120**, 7020-7063 (2020).
11. L. Gaines. Lithium-ion battery recycling processes: Research towards a sustainable course. *Sustain. Mater. & Techno.* **17**, e00068 (2018).
12. *ReCell Center, U.S. Department of Energy (DOE)*, <https://recellcenter.org/>
13. Y. Shi, M. Zhang, Y. S. Meng & Z. Chen. Ambient-Pressure Relithiation of Degraded $Li_xNi_{0.5}Co_{0.2}Mn_{0.3}O_2$ ($0 < x < 1$) via Eutectic Solutions for Direct Regeneration of Lithium-Ion Battery Cathodes. *Adv. Energy Mater.* **9**, 1900454 (2019).
14. P. Xu, Q. Dai, H. Gao, H. Liu, M. Zhang, M. Li, Y. Chen, K. An, Y. S. Meng, P. Liu, Y. Li, J. S. Spangenberger, L. Gaines, J. Lu & Z. Chen. Efficient Direct Recycling of Lithium-Ion Battery Cathodes by Targeted Healing. *Joule* **4**, 2609-2626 (2020).
15. Y. Bi, J. Tao, Y. Wu, L. Li, Y. Xu, E. Hu, B. Wu, J. Hu, C. Wang, J.-G. Zhang, Y. Qi & J. Xiao. Reversible planar gliding and microcracking in a single-crystalline Ni-rich cathode. *Science* **370**, 1313-1317 (2020).
16. J. E. Harlow, X. Ma, J. Li, E. Logan, Y. Liu, N. Zhang, L. Ma, S. L. Glazier, M. M. E. Cormier, M. Genovese, S. Buteau, A. Cameron, J. E. Stark & J. R. Dahn. A Wide Range of Testing Results on an Excellent Lithium-Ion Cell Chemistry to be used as Benchmarks for New Battery Technologies. *J. Electrochem. Soc.* **166**, A3031-A3044 (2019).
17. G. Heppel. *NMC 811 adoption in the auto sector hits a speed bump*, <https://www.crugroup.com/knowledge-and-insights/insights/2018/nmc-811-adoption-in-the-auto-sector-hits-a-speed-bump/> (2018).
18. Z. Xu, M. M. Rahman, L. Mu, Y. Liu & F. Lin. Chemomechanical behaviors of layered cathode materials in alkali metal ion batteries. *J. Mater. Chem. A* **6**, 21859-21884 (2018).
19. H.-H. Ryu, K.-J. Park, C. S. Yoon & Y.-K. Sun. Capacity Fading of Ni-Rich $Li[Ni_xCo_yMn_{1-x-y}]O_2$ ($0.6 \leq x \leq 0.95$) Cathodes for High-Energy-Density Lithium-Ion Batteries: Bulk or Surface Degradation? *Chem. Mater.* **30**, 1155-1163 (2018).
20. W. H. Woodford, W. C. Carter & Y.-M. Chiang. Design criteria for electrochemical shock resistant

battery electrodes. *Energy Environ. Sci.* **5**, 8014-8024 (2012).
21. F. Friedrich, B. Strehle, A. T. S. Freiberg, K. Kleiner, S. J. Day, C. Erk, M. Piana & H. A. Gasteiger. Editors' Choice—Capacity Fading Mechanisms of NCM-811 Cathodes in Lithium-Ion Batteries Studied by X-ray Diffraction and Other Diagnostics. *J. Electrochem. Soc.* **166**, A3760-A3774 (2019).
22. S.-K. Jung, H. Gwon, J. Hong, K.-Y. Park, D.-H. Seo, H. Kim, J. Hyun, W. Yang & K. Kang. Understanding the Degradation Mechanisms of $LiNi_{0.5}Co_{0.2}Mn_{0.3}O_2$ Cathode Material in Lithium Ion Batteries. *Adv. Energy Mater.* **4**, 1300787 (2014).
23. J. Kim, H. Lee, H. Cha, M. Yoon, M. Park & J. Cho. Prospect and Reality of Ni-Rich Cathode for Commercialization. *Adv. Energy Mater.* **8**, 1702028 (2018).
24. A. Van der Ven, J. Bhattacharya & A. A. Belak. Understanding Li Diffusion in Li-Intercalation Compounds. *Acc. Chem. Res.* **46**, 1216-1225 (2013).
25. Y. Shi, G. Chen, F. Liu, X. Yue & Z. Chen. Resolving the Compositional and Structural Defects of Degraded $LiNi_xCo_yMn_zO_2$ Particles to Directly Regenerate High-Performance Lithium-Ion Battery Cathodes. *ACS Energy Lett.* **3**, 1683-1692 (2018).
26. K. Sridharan & T. R. Allen. in *Molten Salts Chemistry* (eds Frédéric Lantelme & Henri Groult) 241-267 (Elsevier, 2013).
27. G. Qian, Y. Zhang, L. Li, R. Zhang, J. Xu, Z. Cheng, S. Xie, H. Wang, Q. Rao, Y. He, Y. Shen, L. Chen, M. Tang & Z.-F. Ma. Single-crystal nickel-rich layered-oxide battery cathode materials: synthesis, electrochemistry, and intra-granular fracture. *Energy Storage Mater.* **27**, 140-149 (2020).
28. H. Li, J. Li, X. Ma & J. R. Dahn. Synthesis of Single Crystal $LiNi_{0.6}Mn_{0.2}Co_{0.2}O_2$ with Enhanced Electrochemical Performance for Lithium Ion Batteries. *J. Electrochem. Soc.* **165**, A1038-A1045 (2018).
29. Y. Liu, J. Harlow & J. Dahn. Microstructural Observations of "Single Crystal" Positive Electrode Materials Before and After Long Term Cycling by Cross-section Scanning Electron Microscopy. *J. Electrochem. Soc.* **167**, 020512 (2020).
30. G. Qian, H. Huang, F. Hou, W. Wang, Y. Wang, J. Lin, S.-J. Lee, H. Yan, Y. S. Chu, P. Pianetta, X. Huang, Z.-F. Ma, L. Li & Y. Liu. Selective dopant segregation modulates mesoscale reaction kinetics in layered transition metal oxide. *Nano Energy* **84**, 105926 (2021).
31. J. Li, A. R. Cameron, H. Li, S. Glazier, D. Xiong, M. Chatzidakis, J. Allen, G. A. Botton & J. R. Dahn. Comparison of Single Crystal and Polycrystalline $LiNi_{0.5}Mn_{0.3}Co_{0.2}O_2$ Positive Electrode Materials for High Voltage Li-Ion Cells. *J. Electrochem. Soc.* **164**, A1534-A1544 (2017).